\documentstyle[12pt,twoside,fleqn,espcrc1]{article}

\newcommand{\AmS}{{\protect\the\textfont2
  A\kern-.1667em\lower.5ex\hbox{M}\kern-.125emS}}

\def\ra{\rightarrow}

\def\be{\begin{equation}}
\def\ee{\end{equation}}
\def\bea{\begin{eqnarray}}
\def\eea{\end{eqnarray}}

\def\calO{{\cal O}}
\def\calM{{\cal M}}

\def\bfq{{\bf q}}

\def\bfr{{\bf r}}

\newcommand{\e}{{\mbox{e}}}

\def\noin{\noindent}
\def\non{\nonumber}
\def\lab{\label}
\def\bib{\bibitem}

\def\be{\begin{equation}}
\def\ee{\end{equation}}
\def\bea {\begin{eqnarray}}
\def\eea {\end{eqnarray}}

\def\dag{\dagger}

\def\ra{\rightarrow}

\def\del{\partial}

\def\mN{m_{\mbox{\tiny N}}}

\def\cL{{\cal L}}

\def\gam{\gamma}

\def\CPT{{\small $\chi$PT}}
\def\cLHB{\cL^{\mbox{\scriptsize HB}}_{\rm ch}}

\def\eg{{\it e.g.\ }}
\def\etal{{\it et al.\ }}

\title{Chiral Symmetry in Nuclei}

\author{Kuniharu Kubodera\address{Department of Physics and Astronomy,
        University of South Carolina, \\
        Columbia, South Carolina 29208, USA}%
        \thanks{e-mail: kubodera@sc.edu. 
          Supported in part by the National Science Foundation, USA, 
          Grant No.PHYS-9602000}}

\begin{document}
\maketitle

\begin{abstract}
Effective field theory is considered 
to provide a highly useful framework
for connecting nuclear physics with the symmetries and
dynamics of the underlying theory of strong interactions, QCD.
Of many issues that are of great current interest in this domain,
I concentrate here on two:
(1) A new class of {\it ab initio} calculations of observables 
in two-nucleon systems;
(2) Attempts to extend chiral perturbation calculations to higher-order terms.
\end{abstract}

\section{Introduction}

One of the major challenges 
in nuclear physics today is to establish
a connection between nuclear dynamics 
and the fundamental QCD.
Effective field theory (EFT) provides 
a natural and useful framework for this purpose.
The basic idea of EFT is simple \cite{wei79,gl84}.
Suppose we are interested in phenomena
characterized by a typical energy-momentum scale $Q$.
We expect that the degrees of freedom 
whose energy scales are significantly larger than $Q$
need not feature explicitly in our Lagrangian.
So, introducing a cut-off scale $\Lambda$
that is reasonably large as compared with $Q$,
we separate our fields (generically denoted by $\Phi$)
into a high-energy part $\Phi_{\mbox{\tiny H}}$
and a low-energy part $\Phi_{\mbox{\tiny L}}$.
Integrating out $\Phi_{\mbox{\tiny H}}$,
we arrive at an effective Lagrangian that only involves 
$\Phi_{\mbox{\tiny L}}$.  
The original Lagrangian 
${\cal L}$ and the effective Lagrangian 
${\cal L}_{{\rm eff}}$ are related as
\be
\int\![d\Phi]\e^{{\rm i}\int d^4x{\cal L}(\Phi)}
= \int\![d\Phi_{\mbox{\tiny L}}]
\e^{{\rm i}\int d^4x{\cal L}_{\rm eff}
(\Phi_{\mbox{\tiny L}})}
\ee
By construction, ${\cal L}_{{\rm eff}}$ inherits 
all the symmetries 
(and the patterns of symmetry breaking, if any)
of the original ${\cal L}$.
Then ${\cal L}_{{\rm eff}}$
is given as the sum of all possible monomials of 
$\Phi_{\mbox{\tiny L}}$ and their derivatives
that are consistent with the symmetry requirements.
Since a term involving $n$ derivatives
scales like $(Q/\Lambda)^n$,
we have perturbative expansion
with respect to $\del_\mu/\Lambda$.
In a nuclear-physics application of EFT,
the original Lagrangian ${\cal L}$ is the QCD Lagrangian,
but, if we are concerned with the energy-momentum regime
$Q\ll \Lambda_{\mbox{\tiny QCD}}\sim 1$ GeV,
the relevant effective degrees of freedoms are not
quarks and gluons but hadrons.
Furthermore, for $Q\leq m_\pi$,
it is reasonable to retain only the pions and nucleons
as effective degrees of freedom.
Meanwhile, chiral symmetry of QCD
must be respected in our effective world.
The resulting effective theory is called 
chiral perturbation theory ($\chi$PT) \cite{gl84,wei90}.
In fact, the inclusion of the nucleon in $\chi$PT
poses a problem because its mass is comparable to 
$\Lambda_{\mbox{\tiny QCD}}$.
The heavy-baryon chiral perturbation formalism
(HB$\chi$PT) allows us to circumvent 
this difficulty \cite{jm91,bkm95}.
$\cLHB$ has, as effective degrees of freedom
only the pions and the large components of nucleons,
and it involves expansion 
in $\del_\mu/\Lambda_{\mbox{\tiny QCD}}$, 
$m_\pi/\Lambda_{\mbox{\tiny QCD}}$
and $\del_\mu/\mN$.
Since $\mN\approx \Lambda_{\mbox{\tiny QCD}} $,
we usually lump together chiral 
and heavy-baryon expansions.
In this combined expansion scheme,
the effective chiral Lagrangian can be organized
in terms of the chiral order index ${\bar \nu}$ 
defined by $\bar{\nu}=d+(n/2)-2$,
where $n$ is the number of 
fermion lines that participate in a vertex,
and $d$ is the number of derivatives
(with $m_\pi$ counted as one derivative). 
The leading order terms are given as \cite{bkm95}
\bea
{\cal L}^{(0)}&= &
 \frac{f^2_\pi}{4} \mbox{Tr} 
[ \partial_\mu U^\dagger \partial^\mu U 
 + m_\pi^2 (U^\dagger +  U - 2) ] \non
\\ &&
 + \bar{B} ( i v \cdot D + g_A^{} S \cdot u ) B 
 - \frac12 \displaystyle \sum_A 
      C_A (\bar{B} \Gamma_A B)^2 
\lab{eq17a}\\
{\cal L}^{(1)} &= &
-\frac{i g_A^{}}{2\mN} \bar{B} 
\{ S \!\cdot\! D, v \!\cdot\! u \} B 
 + 2c_1 m_\pi^2 \bar{B} B \mbox{Tr} 
( U + U^\dagger - 2 ) \non
\\ &&
 + (c_2 \!-\! \frac{g_A^2}{8\mN}) \bar{B} 
(v \!\cdot\! u)^2 B 
 + c_3 \bar{B} u \!\cdot\! u  B  \non
\\ &&
  - \frac{c_9}{2\mN} (\bar{B}B)
(\bar{B} i S \!\cdot\! u B ) 
  - \frac{c_{10}}{2\mN} (\bar{B} S^\mu B) 
(\bar{B} i u_\mu B).
\lab{eq17b}
\eea
Here $U(x)$ is an SU(2) matrix field
related non-linearly to the pion field,
$\xi\equiv\sqrt{U}$,
$u_\mu \equiv i (\xi^\dag \del_\mu \xi
                - \xi \del_\mu \xi^\dag)$, 
$S_\mu=i\gam_5\sigma_{\mu\nu}v^\nu/2$, and 
$D_\mu$ is the covariant derivative acting on the nucleon.
Here we have retained only those terms that
are relevant to our subsequent discussion.
The counting rule of Weinberg \cite{wei90}
is that a Feynman diagram consisting of $A$ nucleons,
$N_E$ external fields,
$L$ loops and $C$-separated pieces
is order of $\calO(Q^\nu)$ with
$\nu = 2 L + 2 (C-1) + 2 - (A+N_E) + \sum_i \bar \nu_i$.

In fact, straightforward chiral counting fails for a nucleus, 
because purely nucleonic intermediate states
that occur in a nucleus
can have very low excitation energies
which invalidate the ordinary chiral counting rule \cite{wei90}.
To avoid this difficulty,
we classify Feynman diagrams into two groups.
Diagrams in which every intermediate state
contains at least one meson in flight
are called irreducible diagrams,
and all others are called reducible diagrams.
The chiral counting should be applied 
only to irreducible diagrams.
The contribution of all the irreducible diagrams 
(up to a specified chiral order)
is then used as an effective operator
acting on the nucleonic Hilbert space.
By summing up infinite series of irreducible diagrams
(either solving the Schr\"odinger equation 
or the Lippman-Schwinger equation),
we take account of reducible diagrams.
This two-step procedure may be referred to as
{\it nuclear chiral perturbation theory} \cite{wei90,kol92}.

\section{Hybrid approach to nuclear \CPT\ }

In applying nuclear \CPT\ to cases that involve 
external probes,
a nuclear transition operator ${\cal T}$ is identified with
a set of all the irreducible diagrams
(up to a given chiral order $\nu$) 
with an external current inserted \cite{rho91}.
In a fully consistent \CPT\ calculation, 
${\cal T}$ is to be sandwiched 
between the initial and final 
nuclear states which are governed 
by the nucleon interactions corresponding to
the $\nu$-th order irreducible diagrams.  
In practice, however, 
we often use initial and final nuclear wavefunctions 
obtained from the Schr\"odinger equation that involves 
phenomenological nucleon-nucleon interactions.
This eclectic treatment may be called 
a {\it hybrid approach} to nuclear \CPT.
The hybrid approach was used extensively
to test the validity of 
the ``chiral filter mechanism" \cite{kdr78,rho91}.
This {\it mechanism} is the statement
that soft-pion exchange (unless suppressed by symmetry or kinematics)
should give a dominant exchange-current contribution.
Rho \cite{rho91} gave a clear interpretation of this dominance 
in the language of chiral counting.
As argued in \cite{kdr78},
the space component of the vector current (${\bf V}$) and
the time component of the axial current ($\!A_0\!$),
in general, are expected to exhibit this dominance.
Regarding the ${\bf V}$,
a detailed calculation based on the hybrid HB$\chi$PT 
was carried out by Park, Min and Rho \cite{pmr95}
for the isovector M1 transition amplitude
in the $n({\rm thermal})+p\ra d+\gamma$ reaction.
The result of their next-to-next-to-leading order calculation
indicates that the soft-pion exchange current indeed
gives a dominant contribution and that,
with the next order corrections added, 
agreement with the data is perfect.
As for the $A_0$,
the enhancement factors 
for first-forbidden ($\Delta J=0$, $\Delta\pi={\rm yes}$)  
$\beta$-decay transitions were calculated
in the hybrid HB$\chi$PT \cite{pmr93}.
The results indicate that 
the bulk of the empirical enhancement factors \cite{war91}
can be attributed to the soft-pion exchange current,
and that the next-order corrections are a small fraction of it.

Furthermore, Park, Min, Rho and myself (PKMR) \cite{pkmr-apj}
have recently computed in HB$\chi$PT
the cross sections for the solar proton burning process
$p+p\rightarrow d+e^++\nu_e$.
The $pp$ fusion rate obtained
in PKMR's calculation
essentially agrees with the rate 
used in the standard solar model \cite{bahcall}.
This result is important in the light of 
Ivanov et al.\ $\!\!$'s recent claim \cite{ivanov98}
that their field theoretic approach gives
a $pp$ fusion rate significantly different than
that obtained from the ordinary nuclear physics approach
based on Schr\"odinger equations.
PKMR's result based on \CPT\ 
does not support the claim in Ref.\cite{ivanov98}.

\section{First attempt at {\em ab initio} calculation}

Despite the impressive success of the hybrid approach,
it is important to formally justify or go beyond 
this approximation.
Very recently we (PKMR) have carried out
a calculation in which the transition operators 
and the nuclear interactions are treated 
on the same footing \cite{pkmr,pkmreff}.
This formally consistent treatment may be called
an {\it ab initio} calculation.
Let me summarize here some salient features of
our work described in \cite{pkmreff}.

We have performed next-to-leading-order (NLO) {\it ab initio}
calculations for the two-nucleon systems
both with and without the pion field.
The purpose of considering these two cases is
to examine whether 
the ``pionless" and ``pionful" effective theories,
which have different cut-off scales,
exhibit behaviors that are 
generally expected for EFT.

Since, to NLO, pion loops do not enter,
we can simply work with a potential.
The {\em bare} potential ${\cal V}$ has the form
\bea
{\cal V}(\bfq) &=&
- \tau_1\cdot\tau_2 \,\frac{g_A^2}{4 f_\pi^2}\,
\frac{ \sigma_1\cdot \bfq\,\sigma_2\cdot \bfq}{\bfq^2+m_\pi^2}
+ \frac{4\pi}{m_{\mbox{\scriptsize N}}}
 \left[C_0 + (C_2 \delta ^{ij} + D_2 \sigma^{ij})
 q^i q^j \right]{\bfq^2} \label{Vq}
\eea
with
$\sigma^{ij} = 3/\sqrt{8}[
(\sigma_1^i \sigma_2^j + \sigma_1^j \sigma_2^i)/2
- (\delta^{ij}/3) \sigma_1 \cdot \sigma_2 ],$
where $\bfq$ is the momentum transfer.
The parameters $C$'s and $D_2$ are defined for 
each isospin channel.
The first (nonlocal) term is the pion exchange 
involving the Goldstone boson and hence completely known
from \CPT.
The (local) terms in the square brackets represents 
the effects of 
the degrees of freedom that have been integrated out.
Use of the potential Eq.(\ref{Vq}) 
in the Lippman-Schwinger equation
generates an infinite series of divergent terms.
How to regularize this divergence
is one of the hottest issues in nuclear \CPT\ 
(see blow).
We use a momentum cut-off scheme \cite{lep97}
and introduce a Gaussian cutoff: 
$V(\bfr) \equiv \int [d^3\bfq/(2\pi)^3]
\, \e^{i\bfq\cdot \bfr}\, S_\Lambda(\bfq^2)\, {\cal V}(\bfq)$,
with  $S_\Lambda(\bfq^2) = \exp( - \bfq^2/2 \Lambda^2 )$.
For a given cutoff
we can determine the constants $C$'s
and $D$'s in (\ref{Vq}) by relating them (after renormalization)
to the scattering length and the effective range 
for the scattering channels
or to a selected set of the deuteron observables.
Then we are in a position to predict
the N-N scattering phase shifts 
and the low-energy properties of the deuteron
(other than those used as input).
Moreover, we can make {\it parameter-free} estimation
of electroweak transition amplitudes.
(In our NLO calculation this is limited to the 1-body 
(impulse-approximation) contributions.)
For this we only need the one-nucleon electroweak current 
derived to leading order.
We have evaluated one-body contributions to the charge
radius $r_d$, the quadrupole moment $Q_d$,
and the magnetic moment $\mu_d$ of the deuteron.
Furthermore, we have computed 
the one-body M1 matrix element 
${\cal M}_{\mbox{\tiny M1}}$ 
for the $np$ capture process
and the Gamow-Teller matrix element
${\cal M}_{\mbox{\tiny GT}}$ 
for the $pp$ fusion process.
The upshot of PKMR's {\it ab initio} calculations is as follows.
(1) All the calculated quantities 
are in good agreement with the empirical information
in both {\it pionless} and {\it pionful} cases.
(2) The results are stable against 
the variation of the cut-off parameter $\Lambda$,
so long as it lies within a reasonable range;
this {\it reasonable} range is found to be
$\Lambda=$100 - 300 MeV 
($\Lambda=$200 - 500 MeV)
in the absence (presence) of the pion,
values that are consistent 
with the general EFT consideration.
(3)The presence of the pion brings a noticeable
improvement in the accuracy of the prediction,
markedly reducing the cutoff dependency.

The above-mentioned ``good agreement"
requires a little more explanation.
First, for the quantities that do not involve
external probes, our calculated values
can be compared directly with the experimental values,
or with the values obtained 
from \eg the Argonne $v18$ potential
(which was constructed to fit the data).
The experimental phase shifts
are reproduced very well
up to $p\approx 70$ MeV without the pion
and up to $p\approx 140$ MeV with the pion.
The deuteron D-state probability
corresponding to the $v18$ potential
is also reproduced satisfactorily.
People may ask, however, what is the difference
between the familiar effective-range formula
and our prediction.
It is true that, in the pionless case,
the two low-energy constants,
$C_0$ and $C_2$ in Eq.(\ref{Vq}),
practically replace the roles of 
the effective-range expansion parameters,
$a$ and $r_e$.
With introduction of the pion, however,
there is no such trivial correspondence,
and \CPT\  expansion contains more physics in it.
Furthermore, \CPT\  enables us to apply
a unified expansion scheme 
not only to N-N scattering
but also to nuclear transition processes.

The quantities that involve electroweak probes
may be classified into two kinds
according to whether two-body contributions
are expected to be very small or rather significant.
To the first kind belong $r_d$, $Q_d$ and $\mu_d$,
while ${\cal M}_{\mbox{\tiny M1}}$
and ${\cal M}_{\mbox{\tiny GT}}$ are of the second kind.
As for the first group, PKMR's results with the
one-body current alone agree well
with the experimental values.
As for ${\cal M}_{\mbox{\tiny M1}}$
and ${\cal M}_{\mbox{\tiny GT}}$, 
PKMR's results are found to be in good agreement 
with the one-body contributions calculated with the use of 
$v18$ potential.
This aspect is of course welcome but precise comparison 
with the experimental ${\cal M}_{\mbox{\tiny M1}}$
can be done only after inclusion of exchange-current contributions.
(For ${\cal M}_{\mbox{\tiny GT}}$ 
there is no experimental data.)
Since a completely self-consistent evaluation of the
exchange-current, which requires an NNLO calculation,
is yet to be done, we may proceed as follows.
We introduce the ratio ${\cal R}$
of the 2-body contribution to the 1-body contribution by
$ \calM_{\mbox{\tiny M1}} =
  \calM_{\mbox{\tiny M1}}^{\mbox{\tiny 1-body}}
+\calM_{\mbox{\tiny M1}}^{\mbox{\tiny 2-body}}
 \equiv \calM_{\mbox{\tiny M1}}^{\mbox{\tiny 1-body}} (1 + {\cal R})$.
Park, Min and Rho \cite{pmr95}
calculated ${\cal R}$ in hybrid HB$\chi$PT 
with the use of the $v18$ potential
and the resulting $\calM_{\mbox{\tiny M1}}$
showed excellent agreement with the experimental value.
Since the ratio ${\cal R}$ is very likely to be much less sensitive
to the wavefunctions than $\calM_{\mbox{\tiny M1}}$ itself,
it is reasonable to expect that ${\cal R}$
calculated in the hybrid approach is close to ${\cal R}$
that would result from an {\it ab initio} evaluation.
This expectation gets additional support from the fact
that the range of the soft-pion exchange current
is about the same as the range probed in our EFT.
In this sense the hybrid approach is justifiable 
from the EFT point of view.

In the above we introduced a momentum cut-off 
to regularize the divergence that appears in
the Lippman-Schwinger equation
with a potential of the type Eq.(\ref{Vq}).
Another regularization method,
called the polynomial divergence subtraction (PDS),
was proposed by Kaplan, Savage and Wise \cite{ksw},
and it has been used extensively to calculate 
various observables in the two-nucleon systems \cite{ksw2,savage}.
One of the advantages of PDS
is that it preserves chiral invariance.
By contrast, the cut-off regularization loses
manifest chiral invariance.
Although this feature is not consequential
in the NLO calculation discussed above,
it does become relevant in higher-order calculations.
Meanwhile, the PDS counting scheme generates
a counter term that is of the same order
as the leading-order pion-exchange current \cite{savage}.
Since the strength of this counter term
needs to be determined using data,
there is significant loss of predictivity.
This is to be contrasted with the cut-off scheme
in which the leading-order pion-exchange contribution
is a prediction of the theory,
and in which the problem of unknown counter terms occurs only
in the loop corrections to the one-pion exchange term \cite{pmr95,pmr93}.
For more discussion on the regularization schemes 
in the two-nucleon systems, see \eg \cite{cohen}.

\section{$p+p \ra p+p+\pi^0$ reaction near threshold}
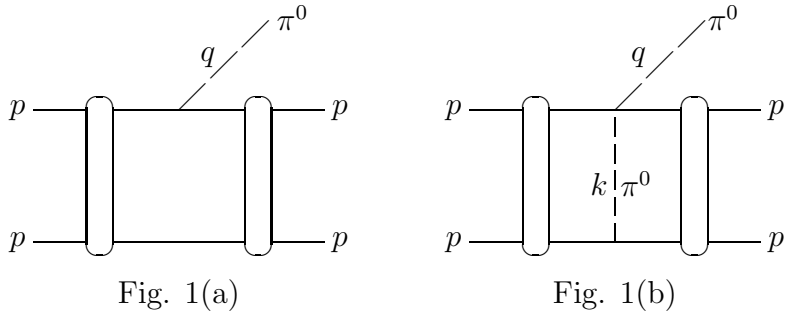
\begin{picture}(400,130)

\put(100,40){\line(1,0){20}}
\put(130,40){\line(1,0){50}}
\put(190,40){\line(1,0){20}}

\put(100,90){\line(1,0){20}}
\put(130,90){\line(1,0){50}}
\put(190,90){\line(1,0){20}}

\put(155,90){\line(1,1){10}}
\put(167,102){\line(1,1){10}}
\put(179,114){\line(1,1){10}}

\put(125,65){\oval(10,60)}
\put(185,65){\oval(10,60)}

\put(166,110){\makebox(0,0){$q$}}

\put(198,125){\makebox(0,0){$\pi^0$}}
\put(94,90){\makebox(0,0){$p$}}
\put(216,90){\makebox(0,0){$p$}}
\put(94,40){\makebox(0,0){$p$}}
\put(216,40){\makebox(0,0){$p$}}

\put(155,20){\makebox(0,0){Fig. 1(a)}}


\put(265,40){\line(1,0){20}}
\put(295,40){\line(1,0){50}}
\put(355,40){\line(1,0){20}}

\put(265,90){\line(1,0){20}}
\put(295,90){\line(1,0){50}}
\put(355,90){\line(1,0){20}}

\put(320,90){\line(1,1){10}}
\put(332,102){\line(1,1){10}}
\put(344,114){\line(1,1){10}}

\put(320,40){\line(0,1){7}}
\put(320,50){\line(0,1){7}}
\put(320,60){\line(0,1){7}}
\put(320,70){\line(0,1){7}}
\put(320,80){\line(0,1){7}}

\put(290,65){\oval(10,60)}
\put(350,65){\oval(10,60)}

\put(329,110){\makebox(0,0){$q$}}

\put(361,125){\makebox(0,0){$\pi^0$}}
\put(259,90){\makebox(0,0){$p$}}
\put(381,90){\makebox(0,0){$p$}}
\put(259,40){\makebox(0,0){$p$}}
\put(381,40){\makebox(0,0){$p$}}
\put(328,62){\makebox(0,0){$\pi^0$}}

\put(314,62){\makebox(0,0){$k$}}

\put(320,20){\makebox(0,0){Fig. 1(b)}}

\end{picture}

\vspace{-0.3cm}
I next discuss the near-threshold
$pp \rightarrow pp\pi^0$ reaction,
which has recently been attracting a great deal of attention
\cite{meyetal90} - \cite{dkms}.
You may wonder what motivates us to study 
this very specific process.
Besides the availability of high-precision data \cite{meyetal90},
a strong motivation comes from the exceptional sensitivity of  
this reaction to higher chiral-order terms.
The corresponding charged-pion production reaction, e.g. 
$pp \rightarrow pn\pi^+$, is described reasonably well 
by the single nucleon process (Born term), Fig.1(a), 
and the ``large" Weinberg-Tomozawa 
s-wave pion rescattering process, Fig.1(b).
By contrast, the Weinberg-Tomozawa term 
does not contribute to the $pp \rightarrow pp\pi^0$ reaction,
rendering it particularly sensitive to and hence 
a good testing ground for the less-well-understood ``small" 
isoscalar s-wave pion rescattering terms.

The first $\chi$PT calculations for $pp \rightarrow pp\pi^0$
were carried out by Park \etal\cite{pmmmk96}
and Cohen \etal\cite{cfmv96}.
In the hybrid HB$\chi$PT they used,
the transition amplitude is given by 
$T\,=\,\langle \Phi_f | \sum_\nu{\cal T}^{(\nu)} | 
\Phi_i \rangle$,
where $|\Phi_i\rangle$ ($|\Phi_f\rangle$)
is the initial (final) two-nucleon state
distorted by the phenomenological N-N interaction.
 ${\cal T}^{(\nu)}$ stands for the transition operator 
of chiral order $\nu$.
The lowest-order one-body impulse term [Fig.1(a)] 
gives ${\cal T}^{(\nu=-1)}$, 
while the lowest-order two-body rescattering term [Fig.1(b)]
gives ${\cal T}^{(\nu=1)}$.
The $\nu=1$ HB$\chi$PT calculations in \cite{pmmmk96,cfmv96}
lead to $pp \rightarrow pp\pi^0$ cross sections 
that are much smaller than the measured values.
Meanwhile, Lee and Riska's work \cite{lr93}
based on the one-boson exchange N-N potential
suggests that shorter range isoscalar meson-exchange processes, 
like $\sigma$- and $\omega$- exchanges 
between the two protons, might be very important 
for the $pp \rightarrow pp\pi^0$ reaction.
To study the behavior of higher chiral-oder terms
and to examine a possible connection between
these higher order terms and the heavy-meson exchange contributions,
it seems of great importance to perform a $\nu=2$ calculation.
Such a calculation has recently been done
by Dmitrasinovic, Myhrer, Sato and myself 
(DKMS) \cite{dkms}. 
I describe here briefly the highlights of our results.

As an initial attempt we may concentrate on
the effective transition operators ${\cal T}^{(\nu)}$ themselves
instead of the full distorted-wave transition amplitude $T$.
Furthermore, we limit ourselves to the {\it threshold kinematics}, 
which means that, in Fig. 1(b),
$q$ = $(m_\pi , \vec{0})$
and $k$ = $(m_\pi/2, \vec{k})$ with $k^2$ = $- m_\pi m_N$. 
Enumerating $\nu$=2 irreducible diagrams
that give rise to transition operator ${\cal T}^{(\nu=2)}$
for $N\!N \rightarrow N\!N \pi$,
we find 20 topologically distinct types of diagrams
(nineteen of them are new). 
For a particular case of the 
$pp \rightarrow pp\pi^0$ reaction near threshold,
the isospin selection rules and the s-wave character 
of the outgoing pion
reduce this number from 20 to 7 (six of them are new).
Some representative diagrams are depicted in Fig.2.

\vspace{0.7cm}
\begin{picture}(400,120)

\put(40,40) {\thicklines \line(0,1){80}}
\put(80,40) {\thicklines \line(0,1){80}}
\put(40,100) {\line(1,0){40}}
\put(40,80){\oval(40,40)[r]}
\put(20,90) {\line(2,-3){20}}

\put(60,20)  {\makebox(0,0){(a)}}

\put(140,40) {\thicklines \line(0,1){80}}
\put(180,40) {\thicklines \line(0,1){80}}
\put(140,80) {\line(2,1){40}}
\put(140,80) {\line(2,-1){40}}
\put(120,110) {\line(2,-3){20}}

\put(170,20){\makebox(0,0){(b)}}

\put(240,40) {\thicklines \line(0,1){80}}
\put(280,40) {\thicklines \line(0,1){80}}
\put(240,70) {\line(2,1){40}}
\put(240,90) {\line(2,-1){40}}
\put(220,110) {\line(1,-1){20}}

\put(260,20){\makebox(0,0){(c)}}

\put(340,40) {\thicklines \line(0,1){80}}
\put(380,40) {\thicklines \line(0,1){80}}
\put(340,80) {\line(1,0){20}}
\put(360,80) {\line(2,1){20}}
\put(360,80) {\line(2,-1){20}}
\put(360,80) {\line(0,1){30}}

\put(360,20){\makebox(0,0){(d)}}

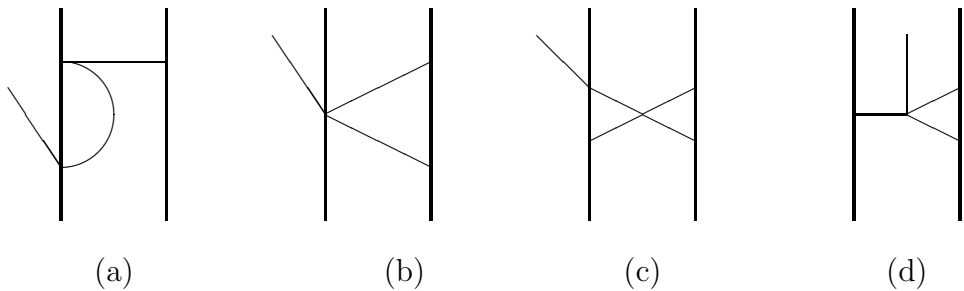
\put(210,0){\makebox(0,0){Fig.2: Selected $\nu=2$ diagrams;
the thick (thin) lines represent nucleons (pions).}}

\end{picture}

\vspace{0.7cm}
\noin
Let ${\cal T}_n^{(\nu=2)}$ ($n=1,2 \ldots, 7$) represent 
the transition operator coming from the $n$-th type of diagrams.
The importance of each operator may be measured
in terms of the ratio
${\cal R}_n\equiv {\cal T}_n^{(\nu=2)}/{\cal T}^{(\nu=1)}$,
where ${\cal T}^{(\nu=1)}$ is
the leading-order two-body transition operator.

Of the seven types of diagrams, 
some can be interpreted as vertex corrections
to the lower order operators,
${\cal T}^{(\nu=-1)}$ and  ${\cal T}^{(\nu=1)}$;
Fig.2(a) gives an example.
The other types are rather loosely called
``two-pion" diagrams;
three examples are shown in Figs. 2(b),(c),(d).

We have found \cite{dkms} 
that ${\cal R}_n$'s corresponding 
to the vertex-correction-type diagrams 
are small ($0.1\sim 0.2$)
in conformity with the general expectation of \CPT.
On the other hand, some of the ``two-pion" diagrams
turn out to give very large contributions.
Especially, ${\cal R}_n$'s 
belonging to the types illustrated
in Fig.2(b), Fig.2(c) and Fig.2(d) 
are individually very large.
For the first two, $|{\cal R}_n|=3\sim 7$
(depending on the input low-energy parameters),
while $|{\cal R}_n|=7\sim 10$ 
for the pion-pion rescattering diagram, Fig.2(d).
This feature is consistent with the expectation
that the $pp\rightarrow pp\pi^0$ reaction
is sensitive to heavy-meson exchanges.  
Thus the phenomenologically important 
$\sigma$-meson contributions \cite{lr93}
seem to have discernible ``representatives"
among the NNLO \CPT\  diagrams.
The large contributions of 
the ``two-pion" diagrams do not necessarily constitute evidence 
for the non-convergence of \CPT\ expansion. 
Since these diagrams appear only at NNLO or higher, 
the convergence can be tested only by calculating corrections 
to the NNLO diagrams.
On the historical note,
I should mention that Gedalin \etal \cite{gmr98} 
considered some of the NNLO diagrams 
and pointed out that they could be large.
According to our calculation,
some important NNLO diagrams
are missing in \cite{gmr98}
and a loop integral expression
in \cite{gmr98} needs to be corrected.

DKMS use the ``standard" chiral counting rule 
of Weinberg \cite{wei90}, whereas
Cohen \etal \cite{cfmv96} have emphasized 
that for the $NN\rightarrow NN\pi$ reaction,
which involves significant energy-momentum transfers,
one should modify the chiral counting rule.
In this new counting scheme,
the expansion parameter is not any longer 
$m_\pi/m_{\mbox{\tiny N}}$
but $\sqrt{m_\pi/m_{\mbox{\tiny N}}}$.
It is a future task 
to perform a calculation similar to that of DKMS
using the modified counting rule.
Furthermore, to obtain transition amplitudes 
that can be directly compared 
with the experimental cross sections, 
it is imperative to carry out 
a distorted-wave calculation.
We hope to be able to report our investigation
along this line in the near future. 

In the first part of my talk
I have surveyed the recent progress 
in \CPT\  treatments of low energy-momentum
transfer observables in the two-nucleon systems.
I have mentioned that we are entering the first  stage 
of {\it ab initio} calculations
based on HB$\chi$PT.
It will be extremely nice if we can perform 
similar {\it ab initio} calculations
for \eg $NN\rightarrow NN\pi$,
which involve higher energy-momentum transfers.
This is a difficult but urgent challenge.

\section*{Acknowledgments}
It is my great pleasure to attend this Symposium
dedicated to Professor Koichi Yazaki, and 
I would like to take this opportunity 
to express my sincere gratitude to Koichi 
for his warm friendship and his kind guidance in physics
for over three decades.
The work described here has been carried out 
in collaboration with Tae-Sun Park, Dong-Pil Min, 
Mannque Rho, Toru Sato, Veljko Dmitrasinovic and Fred Myhrer.
I owe deep thanks to each of these colleagues.

\end{document}